\documentclass[preprint,3p,times,twocolumn,authoryear]{elsarticle}

\usepackage{amssymb}
\usepackage{hyperref}
\hypersetup{colorlinks=true,linkcolor=blue}
\usepackage{multirow}
\journal{Planetary and Space Science}

\begin{document}

\begin{frontmatter}


 \title{Co--orbital resonance with a migrating proto--giant planet}
 \author[l1]{Pablo Lemos\corref{cor1}}
 \ead{plemos@fisica.edu.uy}
 \cortext[cor1]{Corresponding author}
 \author[l1]{Tabar\'e Gallardo}
 \address[l1]{Depto. de Astronom\'ia, Facultad de Ciencias, UdelaR, Montevideo, Uruguay}

\begin{abstract}

In this work we pose the possibility that, at an early stage, the migration of a proto--giant planet caused by the presence of a gaseous circumstellar disk could explain the continuous feeding of small bodies into its orbit. Particularly, we study the probability of capture and permanence in co--orbital resonance of these small bodies, as planets of diverse masses migrate by interaction with the gaseous disk, and the drag induced by this disk dissipates energy from these small objects, making capture more likely. Also, we study the relevance of the circumplanetary disk, a structure formed closely around the planet where the gas density is enhanced, in the process of capture. It is of great interest for us to study the capture of small bodies in 1:1 resonance because it could account for the origin of the Trojan population, which has been proposed \citep{2011Icar..215..669K} as a mechanism of quasi-satellites and irregular satellites capture. 
\end{abstract}


\end{frontmatter}


\section{Introduction}
It is now widely accepted that the planetary systems, and particularly our Solar System, were not formed in an \textit{in--situ} scenario. On the contrary, these planets underwent a substantial migration of their orbits. This migration may have different origins, such as interactions with a gaseous (\citealp{1979ApJ...233..857G}; \citealp{1980ApJ...241..425G}) or planetesimal \citep{1984Icar...58..109F} disk. The presence of these structures imply that the migration must have taken place in the early Solar System. 

Nevertheless, the actual path of each planet still remains as an open question. One of the models explaining this process is the \textit{Jumping--Jupiter} model \citep{2009A&A...507.1041M}, where the young Solar System contains 5 giant planets. These planets migrate through interaction with a planetesimal disk and becomes unstable, and due to this instability, the four giant planets present nowadays have encounters with the fifth one, acquiring orbits similar to the ones we observe in the present and ejecting the remaining one. Besides explaining the orbital characteristics of the giant planets, this model have been successfully applied to explain the origin of many small bodies population in our Solar System (\citealp{2013ApJ...768...45N}; \citealp{2014AJ....148...56B}; \citealp{2014AJ....148...25D}; \citealp{2014ApJ...784...22N}), as well as the Late Heavy Bombardment \citep{2017AJ....153..153D}. On the other side, the constraints given by the observations of the present Solar System impose a large number of conditions to be met by the results of simulations to be considered as successful, which make this model unlikely \citep{2012AJ....144..117N}.

Although a model of capture in co--orbital motion in the frame of the Jumping--Jupiter scenario (i.e., involving the encounter of Jupiter with an ice giant) has been proposed \citep{2013ApJ...768...45N}, we pose the hypothesis that a population of planetesimals could be captured in the co--orbital region of the migrating planet before the instability, when the capture could be assisted by the drag force exerted on the planetesimals by the gaseous disk. In a subsequent stage, the particles captured as Trojans can transform their orbits into quasi-satellites, which in turn enables them to have close encounters with the host planet \citep{2011Icar..215..669K}.

In this work we investigate two different stages of the capture process. Firstly, we analyse the probability of capture in the planet's co--orbital region. A planetesimal is said to be captured if it enters and stays in the co--orbital region for a certain period of time, which implies a change in semi--major axis of the bodies. It is worth noting that, in this scenario, this drift is produced by two different mechanisms: for the planet, it is caused by the gravitational interaction with the disk, while the aerodynamical gas drag is the reason for the migration of planetesimals. The different migration timescales involved can account for the continuous feeding of bodies in the mentioned region, where they can be captured in resonance. In second place, we study the survivability of planetesimals in this region once they are captured.

The outline of this study is as follows: In Section \ref{sec:met} we present the hypothesis and model setup used for this work. In Section \ref{sec:cap} we analyse the results of the simulations made, looking for captures in co--orbital resonance and their dependence on the presence of a gas disk along with the migration timescale, and study the survivability of these bodies. Finally, we present the conclusions, discussion and future work in Section \ref{sec:con}.

\section{Methods}
\label{sec:met}

\subsection{Disk parameters}

In this work we study 3 different cases for the gas drag: Experiment A has no presence of any gas, so the forces acting on the planetesimals are only of gravitational origin. We use this prescription as a control run, to test the effect of the gas. Experiments B and C use a circumstellar (CS) disk of gas where the radial distribution of surface density is

\begin{equation}
\Sigma_s(r)=1700\,\left(\frac{r}{1\,ua}\right)^{-3/2}g\,cm^{-2}
\label{eq:gas}
\end{equation}

which corresponds to the Minimum Mass Solar Nebula \citep{1977MNRAS.180...57W}. The vertical distribution of the volume density is given by

\begin{equation}
	\rho(z)=\rho_0e^{-z^2/2h^2}
\end{equation}

where $\rho_0=\rho(z=0)$ is the mid--plane density and $h$ is the vertical scale height. When a vertically isothermal disk is assumed, $h=c_s/\Omega$, where $c_s$ is the sound speed and $\Omega=\sqrt{GM_{\odot}/r^3}$ is the Keplerian angular velocity, the relation between surface and volume density is

\begin{equation}
	\rho_0=\frac{1}{\sqrt{2\pi}}\frac{\Sigma_s}{h}
\end{equation}

Besides this circumstellar disk, Experiment C uses a circumplanetary (CP) disk following \citet{2013AJ....146..140F}. In that work, the density and sound speed of the circumplanetary disk are parametrized as

\begin{equation}
	\Sigma_p=\Sigma_d \left(\frac{r}{r_d}\right)^{-p}\;\;,\;\;\;\;\; c_s=c_d \left(\frac{r}{r_d}\right)^{-q/2}
	\label{eq:par}
\end{equation}

where $r_d\equiv d r_H$ is a typical length scale related to the Hill radius $r_H$, approximately equal to the effective size of the circumplanetary disk, and $c_d$ is the sound speed at this point. In this work we set $p=3/2$, $d=0.4$ and $q=1/2$ based on the result of previous hydrodynamical simulations \citep{2008ApJ...685.1220M}. Lastly, we choose $\Sigma_d=1.0\,g\,cm^{-2}$ following \citet{2013AJ....146..140F}. The inclusion of this circumplanetary disk is important because the planetesimals are expected to enter in the co--orbital region nearby the embryo, and the enhanced energy dissipation provided by this structure could modify substantially the dynamics of these bodies.

Lastly, we dealt with the gas velocity. Due to the gas pressure support, the velocity of the gas around the central star is not Keplerian. Using the parametrization given in \ref{eq:par}, the gas velocity in Experiments B and C can be written as

\begin{equation}
	v_g=v_K(1-\eta)
\end{equation}

where $v_K$ is the Keplerian velocity around the central star or embryo in Experiments B or C respectively, and $\eta$ is given by \citep{2002ApJ...565.1257T}

\begin{equation}
	\eta=\frac{1}{2}\frac{h^2}{r^2}\left(p+\frac{q+3}{2}+\frac{q}{2}\frac{z^2}{h^2}\right)
\end{equation}

The solid bodies that reach a radius $s\simeq 1\,m$ become less coupled with the gas, orbiting the central star with a Keplerian velocity, to the first approximation. This means that there exist a difference between the velocities of gas and rocks, which leads to a friction force, called \textit{gas drag}. The gas drag force for a spherical particle of radius $s$ is given by

\begin{equation}
	F_g=- \frac{1}{2} C_D\,\pi s^2\,\rho \Delta v^2
	\label{eq:gasdrag}
\end{equation}

where dimensionless parameter $C_D$ is the \textit{drag coefficient}, $\rho$ is the gas density and $\Delta v$ is the relative velocity of the particle with respect to the gas.

\subsection{Orbital dynamics simulations}

For this work we use the \textit{Mercury} package \citep{1999MNRAS.304..793C}. The system is composed by a giant planet embryo ($M_P=10M_{\oplus}$) in a circular orbit with semi--major axis of $8\,au$, and migrates with a constant rate of $1.5\times 10^{-5}\,au\,yr^{-1}$, consistent with the rate found by \citet{2002ApJ...565.1257T}. This migration rate is achieved by imposing an acceleration $\vec{a}_M$ of the form 

\begin{equation}
\vec{a}_M=\frac{\dot{a}}{2a^2} \sqrt{\frac{GM_{\odot}}{2/r-1/a}} \hat{v}
\label{eq:tasa}
\end{equation}

acting only on the embryo. Although this force does not represent any actual physical effect, it generates the desired migration rate with minimal impact over the remaining orbital elements. It is important to note that in this work we neglect the embryo's mass growth. The linear model used by \citet{2002ApJ...565.1257T} to derive the migration rate for the embryo assumes an isothermal equation of state, not only vertically but also in the radial direction. Nevertheless, this model has numerous caveats (disks are probably turbulent and non--isothermal, there exist considerable uncertainties in the formulas used to compute torques exerted by the disk, etc.) which 	could introduce variations in the rates found. For this reason we add an ad--hoc constant coefficient $k$ to Equation \ref{eq:tasa} to investigate the relevance of the migration timescale on our scenario in all the simulations made, where the values adopted for $k$ were $1$, $0.1$ and $1.5$.

We generate a swarm of 1000 planetesimals within a ring between $6$ and $10\,au$ with eccentricities and inclinations randomly chosen with a uniform distribution in the interval $\left[0, 0.1\right]$ and $\left[0^{\circ}, 10^{\circ}\right]$ respectively. For the purpose of calculation of the gas drag force given in Equation \ref{eq:gasdrag}, $C_D$ was chosen to be equal to unity, and we assumed a spherical shape for the planetesimals, with radius uniformly distributed between $50\,m$ and $35\,km$, and a uniform density of $1\,g/cm^3$. Using these prescriptions, the total mass of the planetesimal disk is approximately $2.2\times 10^{-11}M_{\odot}$. 

We integrate the system for $10^5$ years using the Bulirsch-Stoer algorithm \citep{stoer2002introduction}, along with user-defined forces to take account of both migration and gas drag. Although this is the slowest algorithm implemented on Mercury, is especially suitable for dealing with close encounters, which is fundamental for this work.

\section{Captures and survivability}
\label{sec:cap}
\subsection{Co--orbital capture}

\begin{figure}
\begin{center}
	\includegraphics[width=\linewidth]{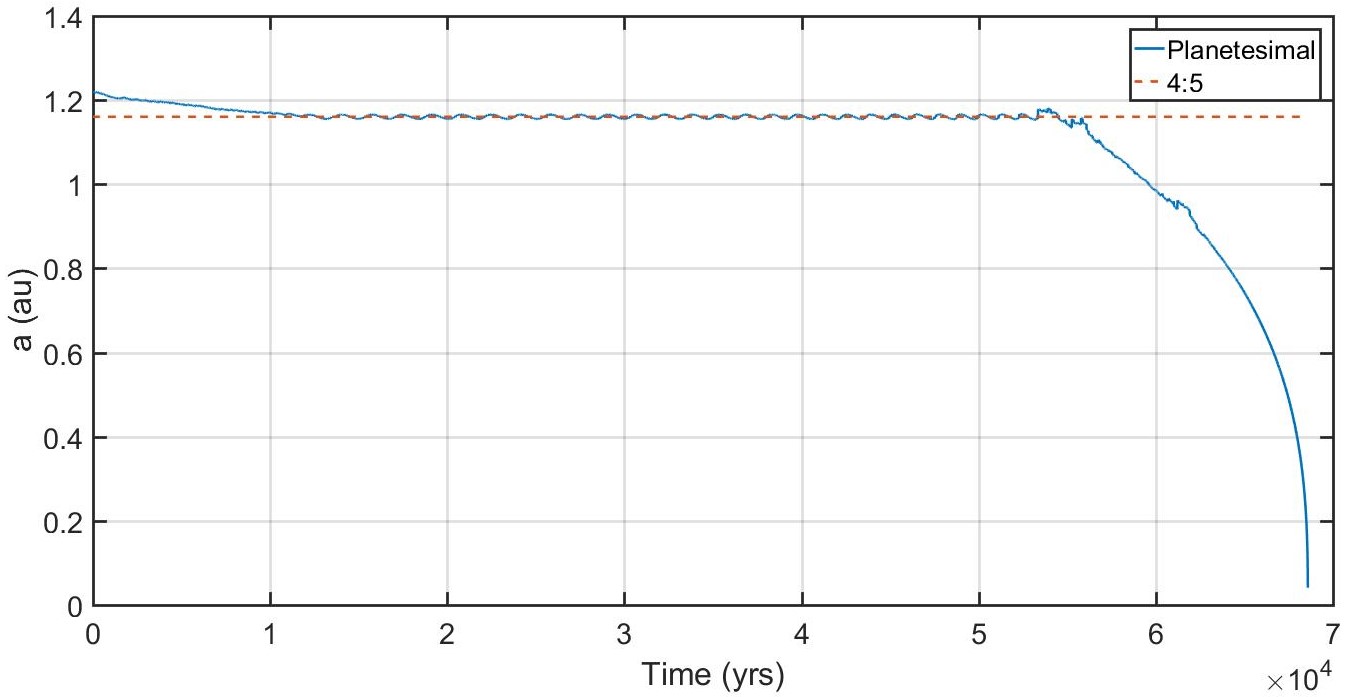}
	\caption{Example of an external body that gets captured in the 4:5 mean motion resonance with the embryo in Experiment C. The capture inhibits the body from crossing the orbit of the embryo for $4\times 10^5$ years, when it escapes and falls to the central star.}
	\label{fig:reso}
\end{center}
\end{figure}

The first condition to obtain bodies captured in co--orbital resonance is that the orbits of the planetesimals cross the orbit of the embryo. If the semi--major axis of a planetesimal is initially larger (smaller) than the embryo's, and in some moment of its evolution become smaller (larger), the body is said to have a crossing orbit. In Experiment A we expect to find a greater number of planetesimals with orbits initially internal to the embryo to become crossing bodies because only the latter has an external force modifying its orbit. However, in Experiments B and C we expect to find the opposite, as the planetesimals migration timescale due to the gas drag is shorter. This different migration rates generate a rapid variation in the semi--major axis ratio between planetesimals and embryo, which implies in addition to a possible crossing orbit, a mean motion resonance capture. For this reason we also look for bodies captured in mean motion resonances with the embryo for at least 1000 years. If the resonances are identified as $|p+q|:|p|$, where $p$ is the degree and $q>0$ the order, $p>0$ implies that the planetesimal has an orbit internal to the embryo and $p<0$ the opposite. We look for all the possible resonance captures with $p\in [-11,10]$ and $q\leq 10$. The next condition analysed is the co--orbital capture. For a capture in co--orbital resonance to take place, we impose the condition that the semi--major axis of the captured body must remain in the co--orbital region of half--width the Hill radius for at least 1000 years. Lastly, we look for bodies that collide with the embryo. It is important to notice that a single body can belong to more than one of the categories listed above. We summarize the results of our simulations in Table \ref{tab:res}.

\begin{table*}[t]
	\centering
	\begin{tabular}{c  c  c  c  c  c  c  c  c  c}
	Experiment & Migration coefficient $k$ & \multicolumn{2}{c}{Crossing} & \multicolumn{2}{c}{MMR capture} & \multicolumn{2}{c}{Co--orbital capture} & \multicolumn{2}{c}{Collision}\\
	&  & \textit{Int} & \textit{Ext} & \textit{Int} & \textit{Ext} & \textit{Int} & \textit{Ext} & \textit{Int} & \textit{Ext} \\
	\hline\hline
	\multirow{3}{*}{A} & 1 & 93 & 34 & 64 & 13 & 3.9 (4.2) & 5.5 (16) & 0.4 & 0\\
& 0.1 & 37 & 39 & 54 & 53 & 0.6 (1.6) & 7.3 (19) & 0 & 0.2 \\ 
No gas disk & 1.5 & 100 & 33 & 57 & 8.1 & 4.9 (4.9) & 3.7 (11) & 0 & 1.2 \medskip\\
	\multirow{3}{*}{B} & 1 & 18 & 100 & 1.4 & 13 & 0.4 (2.3) & 10 (10) & 0 & 0\\
& 0.1 & 17 & 100 & 0.6 & 12 & 0.4 (2.4) & 7.7 (7.7) & 0 & 0.2 \\ 
CS disk & 1.5 & 18 & 100 & 1.0 & 15 & 0.4 (2.3) & 8.4 (8.4) & 0 & 0\medskip\\
	\multirow{3}{*}{C} & 1 & 18 & 100 & 2.2 & 12 & 1.2 (7.0) & 7.3 (7.3) & 0 & 0\\
& 0.1 & 17 & 100 & 1.0 & 12 & 0.4 (2.4) & 8.6 (8.6) & 0 & 0.2 \\ 
CS+CP disk & 1.5 & 18 & 100 & 1.4 & 16 & 0.6 (3.5) & 9.6 (9.6) & 0 & 0 \\
	\end{tabular}
	\caption{Percentage of the total number of bodies in each category for all the simulations including a migrating embryo ($10M_{\oplus}$). The same percentage but referenced only to the crossing bodies is shown in brackets. As expected, the presence of the circumstellar disk generate an increase in the external crossing bodies.}
	\label{tab:res}
\end{table*}

\begin{figure}
\begin{center}
	\includegraphics[width=\linewidth]{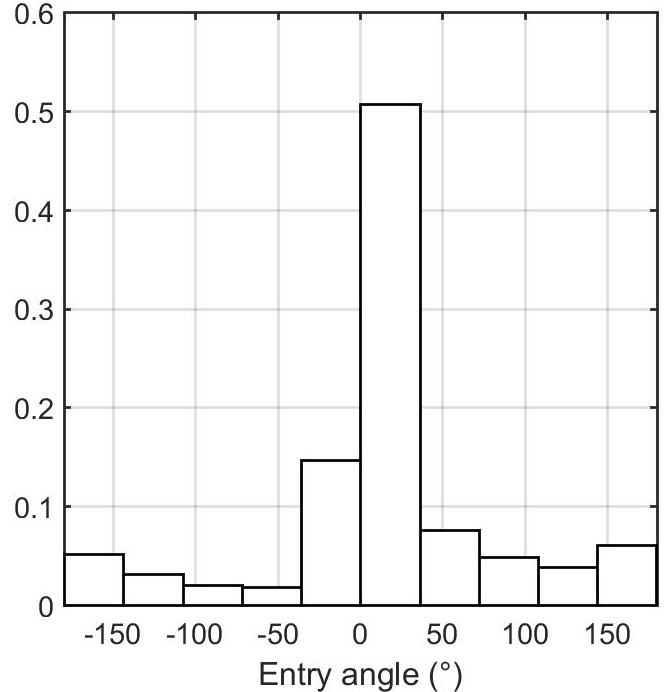}
	\caption{Entry angle for the planetesimals in Experiment C with $k=1$. This trend is the same in all the cases studied. The bodies with entry angle near zero are more susceptible to be affected by the presence by the circumplanetary disk.}
	\label{fig:angent}
\end{center}
\end{figure}

As expected, the presence of the circumstellar disk in experiments B and C generate a greater number of planetesimals in external orbits to become crossing bodies, and also a decrease in the number of internal ones in these categories. When we focus on the bodies trapped in mean motion resonances, we find in first place that they can exist in all the studied cases (Figure \ref{fig:reso}), but the gas drag generates a perturbation too strong for the planetesimals to become easily captured in mean motion resonances, as in Experiment A. In Experiments B and C, the MMR capture probabilities are similar, but the presence of the circumplanetary disk increase the number of planetesimals captured in mean motion resonances with nominal positions closer to the planet (e.g. 9:11, $a_{res}/a_E=1.14314$) than the ones in Experiment B. The relevance of the circumplanetary disk studied in Experiment C is also confirmed if we analyse the entry angle, defined as the planetesimal-central star-embryo angle (i.e. the synodic angle) at the entrance in the co--orbital region. By doing so, an accumulation of bodies entering with angles near zero can be noted (Figure \ref{fig:angent}); these bodies are more susceptible to be affected by the presence of the circumplanetary disk. 

On average, the 1:1 capture probability is $2.3\%$ and $4.3\%$ for internal, and $8.7\%$ and $8.5\%$ for external bodies, referred to Experiments B and C respectively. This result is greater than the estimates given in previous works \citep{2017arXiv170400550N}, taking into account the low inclinations of the planetesimals. Also, the presence of both gaseous disks enhance the total co--orbital capture probability compared with the case with only the circumstellar one, except in the Experiment C with $k=1$ for the external bodies. We try to give an explanation for this phenomenon in the next Section.

Finally, in all the cases studied we find that the collision probability is low. This effect could be explained stating that the mass of the embryo is too low to deflect a significant number of planetesimals into collision orbits, even if assisted by the circumplanetary disk.

\subsection{Survivability}
\label{sub:sur}

\begin{figure}
\begin{center}
	\includegraphics[width=1\linewidth]{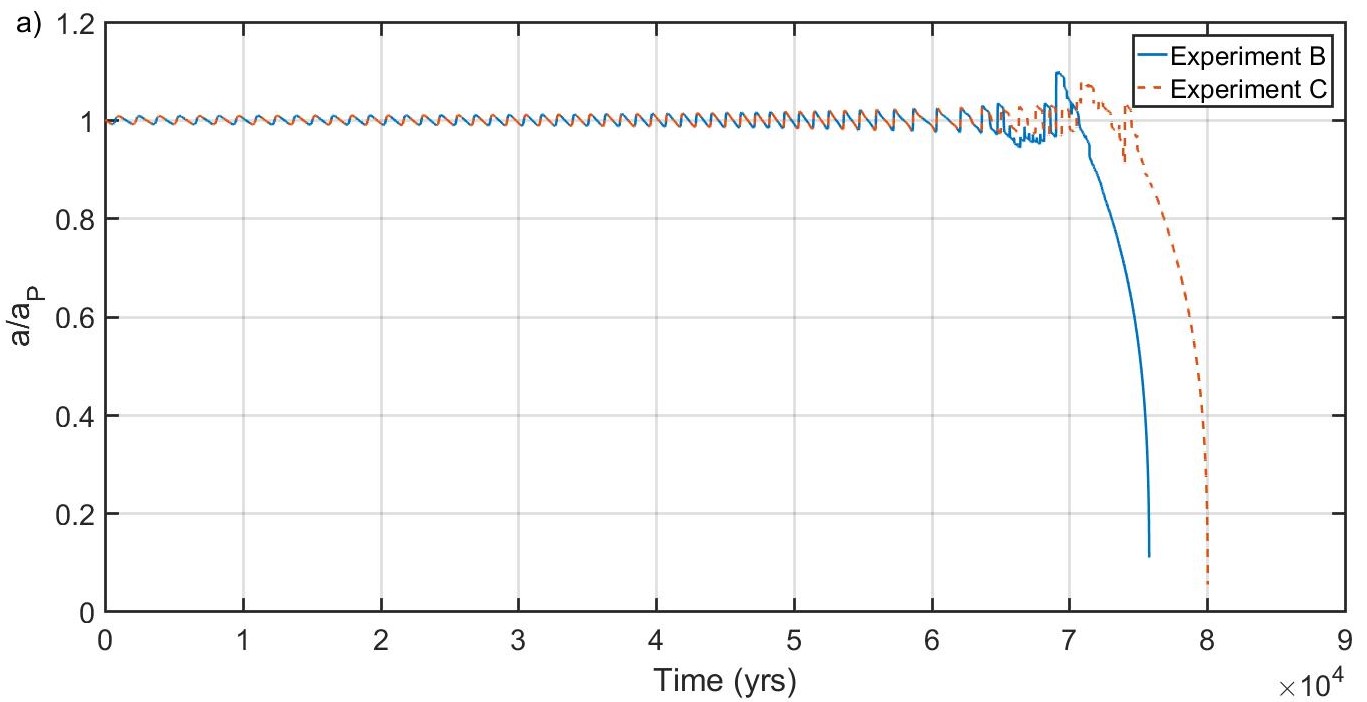}\label{fig:sobrea}
	\includegraphics[width=1\linewidth]{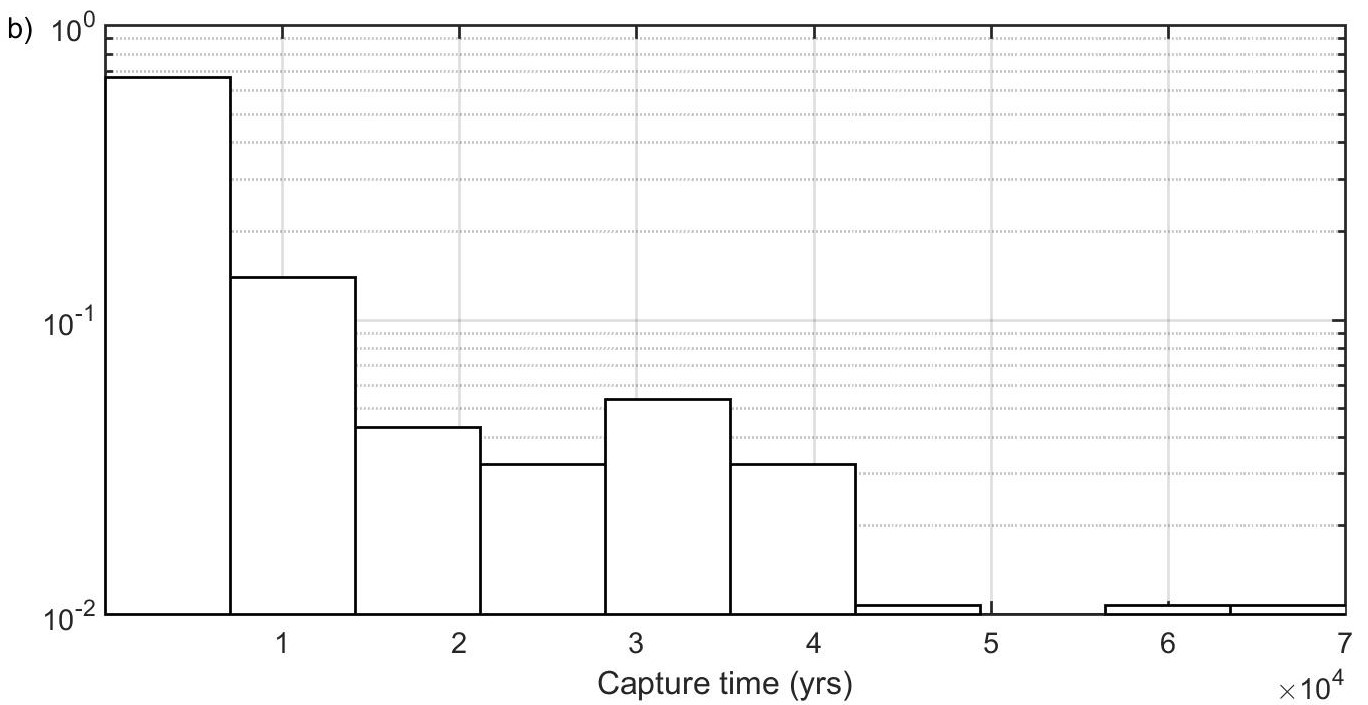}\label{fig:sobreb}
	\caption{\textbf{a)} Lifetime of a particle in Experiment B disk compared with the same particle in Experiment C. The circumplanetary disk can account for the small increase in capture time. \textbf{b)} Histogram of capture time for planetesimals starting the integration inside the coorbital region of the embryo for Experiment C. There is no significant difference between this result and the one for Experiment B.}
	\label{fig:sobre}
	\end{center}
\end{figure}

\begin{figure}
\begin{center}
	\includegraphics[width=1\linewidth]{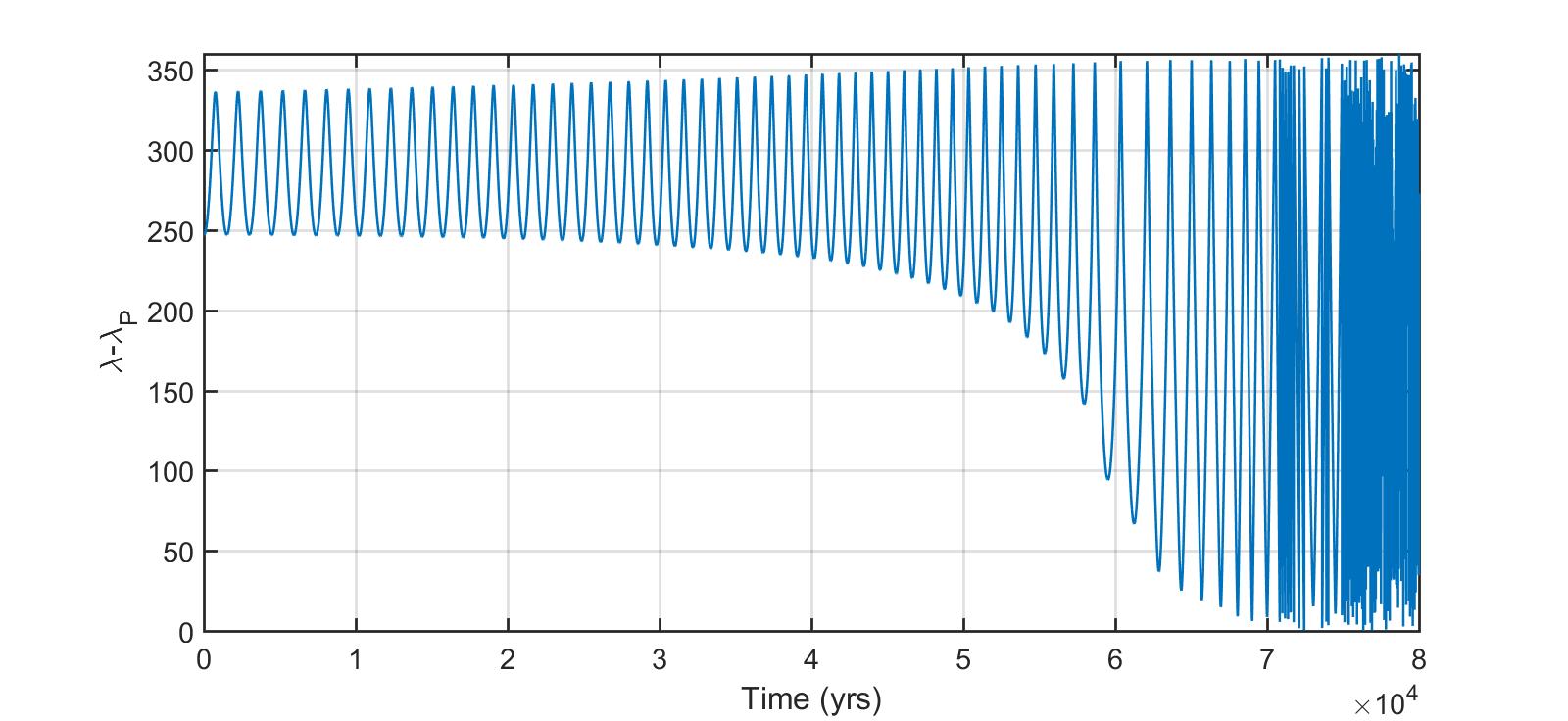}
	\caption{Example of evolution of the critical angle $\lambda-\lambda_P$ for a particle in Experiment C initially placed inside the co--orbital zone. The transition between the Trojan and horseshoe orbit can be seen at $\sim 60000$ years.}
	\label{fig:horse}
	\end{center}
\end{figure}

For this part we work with two different scenarios. In the first one, we place a small number ($\sim 100$) of bodies in the co--orbital region of the embryo to check if these bodies can survive in this state. These results are shown in Table \ref{tab:surv}. From these results two important points should be noted. In first place, the mean capture time is reduced by roughly one order of magnitude when the circumstellar disk is present, and in second place the presence of the circumplanetary disk does not affect the mean or maximum capture time in a substantial way. Nevertheless, the presence of this structure can modify the capture time for certain bodies (Figure \ref{fig:sobre}). The main point to explain this effect is that the planetesimals don't remain all the capture time as Trojans, but in horseshoe orbits, as can be seen in Figure \ref{fig:horse}. This is very important to explain the effect of the circumplanetary disk: as the body become less bounded to the 1:1 resonance, the angle $\lambda-\lambda_P$ increase its libration amplitude, making the planetesimal to approach the embryo. At some point, the planetesimal enters in the zone of the circumplanetary disk, where the gas drag force is greater and reduces more effectively the relative velocity of the planetesimal respect to the embryo.

\begin{table*}[t]
	\centering
	\begin{tabular}{c  c  c  c}
	Experiment & Migration rate & Mean capture time (yrs) & Max. capture time (yrs)\\
	\hline\hline
	\multirow{3}{*}{A} & 1 & 52507 & 100000\\
& 0.1 & 57972 & 100000\\ 
No gas disk& 1.5 & 51439 & 100000\medskip\\
	\multirow{3}{*}{B} & 1 & 6159 & 64820\\
& 0.1 & 6424 & 86230\\ 
CS disk& 1.5 & 6038 & 64300\medskip\\
	\multirow{3}{*}{C} & 1 & 6271 & 70550\\
& 0.1 & 6257 & 74020\\ 
CS+CP disk& 1.5 & 6032 & 67140\\
	\end{tabular}
	\caption{Mean and maximum time of maintenance in the co--orbital region for all the cases studied. The presence of the gaseous disk lower both the mean and maximum time of capture, while a lower migration rate for the embryo helps to prevent the loss of the captured bodies.}
	\label{tab:surv}
\end{table*}

As seen in the previous Section, the mechanism proposed has a non negligible probability of capturing bodies in co--orbital resonance and maintain them for a considerable period of time, even without taking into account the contribution of the mass growth of the embryo due to gas accretion. The effect of mass growth has been already proposed as a possible mechanism to explain the capture of Trojans \citep{1998A&A...339..278M}, although it has some serious drawbacks when explaining their eccentricities and inclinations, for instance \citep{2014SoSyR..48..139S}. In the scenario studied in this article, the mass of the embryo is close to the estimated critical mass needed to trigger the runaway gas accretion phase. For this reason the mass of the planet is expected to increase roughly one order of magnitude in a short period of time, which may lead the planetesimals that were already captured to become more bound to the co--orbital resonance.

Taking this process into account, we now investigate the same scenario but with a migrating giant planet with $1\,M_J$ and initial semi--major axis $a=8\,ua$ instead of the embryo. This change implies that the migration timescale must be much shorter ($\dot{a}=5.3\times 10^{-5} au\,yr^{-1}$) \citep{2003MNRAS.341..213B}, and so is the integration time, which lasts 75000 years in this case. In addition to this, the migration is of Type II, so we impose a gap in the co--orbital region of the planet where the gas density is 100 times smaller than the expected value given in Equation \ref{eq:gas}. The gap half-width $\Delta_{gap}$ is equal to the Hill radius of the planet, which yields a relative half-width $\frac{\Delta_{gap}}{a}\simeq 0.069$. Our value is consistent with the ones obtained in both 2D \citep{2016PASJ...68...43K} ($\frac{\Delta_{gap}}{a}\simeq 0.065$) and 3D simulations \citep{2016ApJ...832..105F} ($\frac{\Delta_{gap}}{a}\simeq 0.1$). We complete the experiment placing 1000 planetesimals inside the planet's co--orbital region, randomly distributed with the same characteristics as the previous integrations.

\begin{figure}
\begin{center}
	\includegraphics[width=1\linewidth]{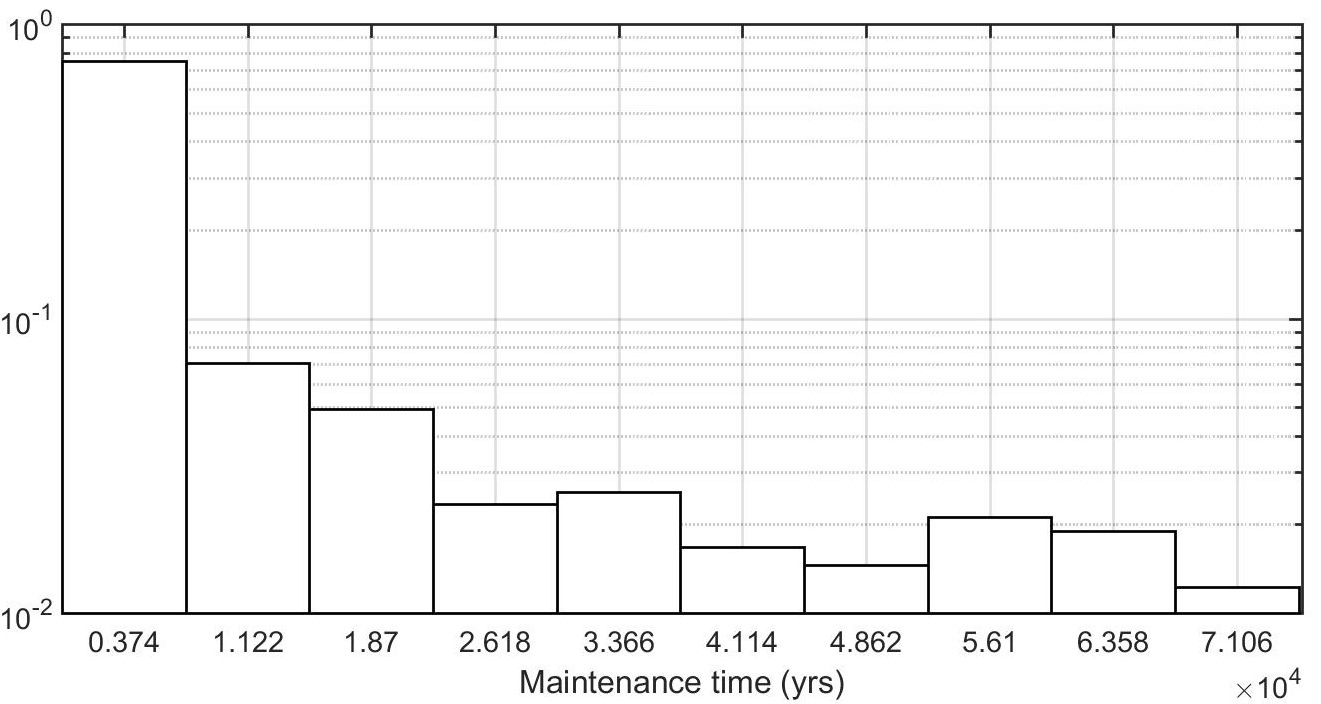}
	\caption{Maintenance time of particles in the co--orbital region for a migrating giant planet. Comparing with Figure \ref{fig:sobre}b, we find that the trend is similar to the previous Experiments.}
	\label{fig:histJ}
	\end{center}
\end{figure}

As we see in Figure \ref{fig:histJ}, the trend in the bodies permanence in co--orbital motion is similar to the previous cases studied, with most of the bodies lost from the co--orbital region in the beginning of the integration. However, and despite of the faster migration of the planet, the percentage of bodies that neither collide with the planet nor the central star is $95.7\%$. $71.4\%$ of these bodies weren't on the co--orbital region at the end of the integration, but the remaining lasted all integration time captured as co--orbitals, so we conclude that $27.4\%$ of all the simulated bodies survive the entire time in a co--orbital resonance with the migrating giant planet.

The asymmetry between populations in $L_4$ and $L_5$ for the surviving bodies is also analysed. This is an important point when comparing with the factual trojan population (see \citealp{2002aste.book..725M} for a review). Defining the temporarily captured objects (TCO) as the population that remain in the co--orbital region for at least 1000 years ($70\%$ of the total), there are three possibilities for their critical angle: librate around $60°$ in the $L_4$ point, librate around $300°$ in the $L_5$ point and oscillate with an amplitude $\sim 300°$ in a horseshoe orbit. Defining $f_{45}\equiv N(L_4)/N(L_5)$, we find that there exist an asymmetry in the number of bodies librating around these locations, obtaining $f_{45}=0.97$. Although the difference between populations in both points is small, there is a significant discrepancy with the parameter predicted using observations of the Solar System, where the expected value is $f_{45}=1.2-1.8$ (\citealp{2013ApJ...768...45N} and the references therein). Additionally, if we analyse further these results we find that this asymmetry between leading and trailing points can be also detected using other tracers, such as the survival probability: defining $p=\frac{N_f}{N_i}$, with $N_i$ the initial number of TCO and $N_f$ the number of TCO at end of simulation, we find that $p(L_4)=54.8\%$ and $p(L_5)=56.9\%$. Also, we find that the mean maintenance time of these bodies is slightly different, with 57009 years for bodies in $L_5$ and 54901 for the ones in $L_4$.

The remaining probability is the horseshoe orbit. From the TCO, $30\%$ have orbits of this type, but the mean maintenance time for them is 8870 years.

We now focus on the final fate of the planetesimals. There are five different possibilities for them: collide with the planet, collide with the central star, ejected from the system, survive outside the co--orbital region (diffusion) or remain captured. We find that the bodies which escape from the co--orbital region have encounters with the planet prior to it, while the bodies that remain inside for all the integration time don't. It is important to note that most of the bodies having encounters with the planet are pushed to the outer parts of the system, where they can remain for a long time due the low density of the gas in this zone. These results are shown on Table \ref{tab:gig}.

\begin{table}[t]
	\centering
	\begin{tabular}{c  c}
	Fate & Percentage of bodies\\
	\hline\hline
	Co--orbitals & 27.2\\
	Collision w/planet & 2.4\\
	Collision w/central star & 0.2\\
	Ejected & 2.3\\
	Diffusion & 67.9\\
	\end{tabular}
	\caption{Fates of the bodies initially placed inside the co--orbital region of a migrating giant planet ($M=1M_J$).}
	\label{tab:gig}
\end{table}

\subsection{Size distribution}

\begin{figure}
\begin{center}
	\includegraphics[width=1\linewidth]{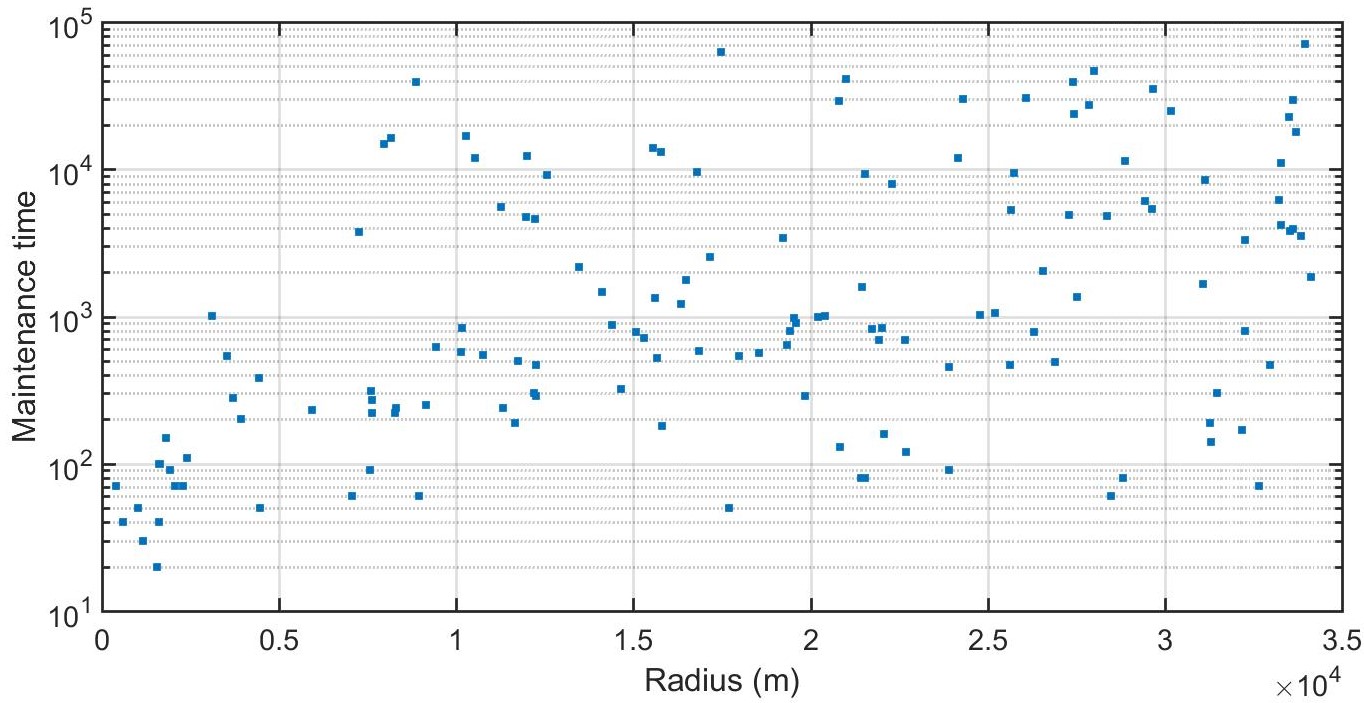}
	\caption{Maintenance time as a function of the particle radius in the case with a migrating embryo. A direct proportionality between $r$ and $\log(t)$ can be seen up to $r\sim 5$ km, where the distribution becomes uniform.}
	\label{fig:tamsobreem}
\end{center}
\end{figure}

From the objects initially located inside the embryo's co--orbital zone, we place our attention in the maintenance time as a function of the radius of the planetesimal (Figure \ref{fig:tamsobreem}). We find that, for radius less than 5 kilometres, the variables are roughly proportional. From this value onwards, the distribution is roughly uniform.

\begin{figure}
\begin{center}
	\includegraphics[width=1\linewidth]{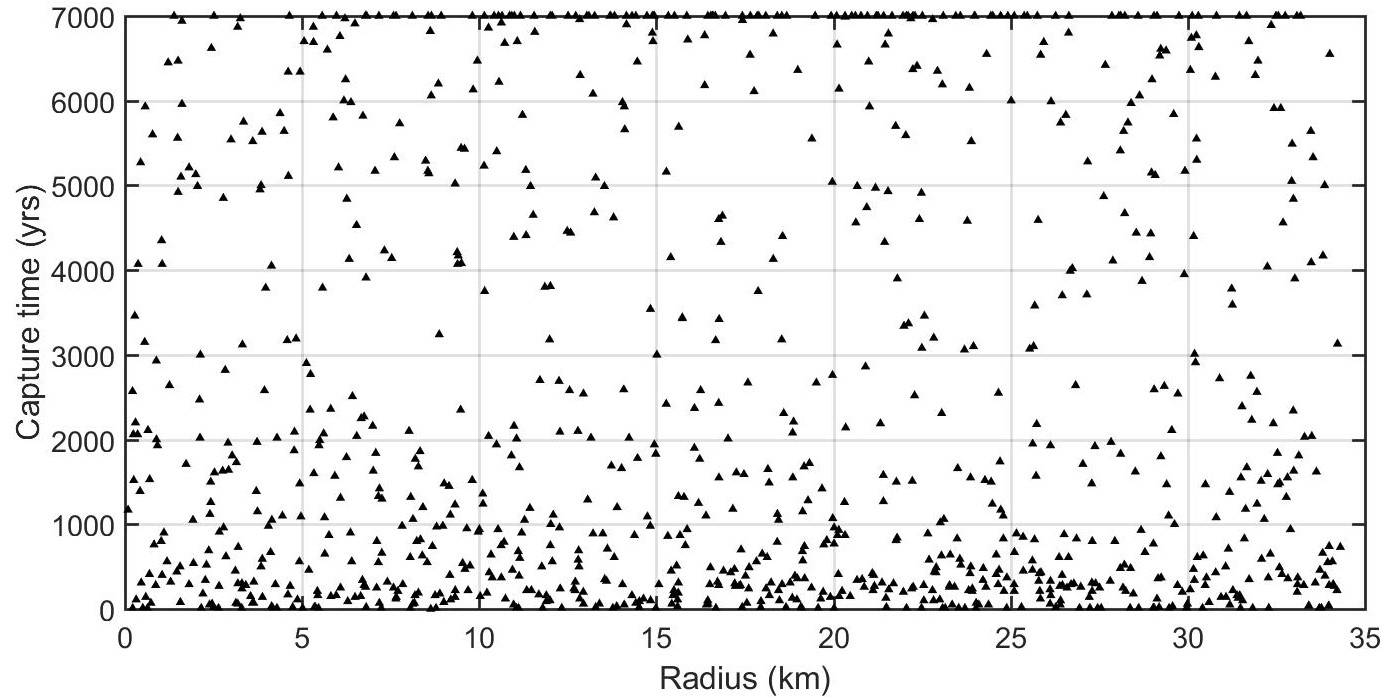}
	\caption{Maintenance time as a function of the particle radius in the case with a migrating giant planet. The smaller objects, along with the ones with greater libration amplitudes (see text), have shorter lifetimes in the co--orbital region.}
	\label{fig:captime}
\end{center}
\end{figure}

Next, we focus on studying the size distribution of the objects in the co--orbital region of the migrating giant planet as a function of its maintenance time (Figure \ref{fig:captime}). At first sight, an accumulation of bodies with short capture times can be seen. This is expected for small objects, for which the drag force is more relevant, but a significant number of bigger bodies also have short capture times. This corresponds to objects with initial conditions far from $L_4$ and $L_5$ points, for which the libration amplitude is big, and therefore are less bound to resonance.

\begin{figure}
\begin{center}
	\includegraphics[width=1\linewidth]{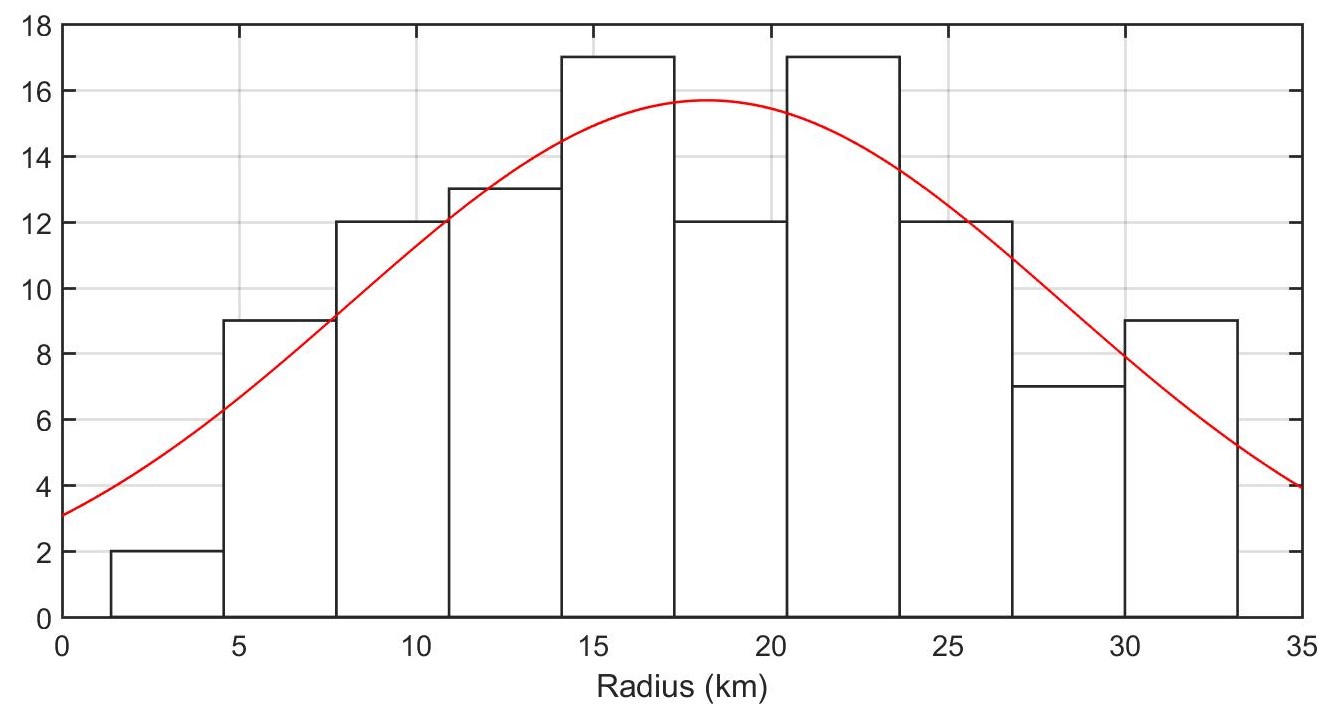}
	\caption{Radius of the particles surviving all the integration time in the co--orbital region for the case with a migrating giant planet, along with a Gaussian fit.}
	\label{fig:tamsobre}
\end{center}
\end{figure}

Lastly, if the focus is placed on the surviving objects, there can be noted from Figure \ref{fig:captime} that the distribution is not uniform. For this reason we plot the radius of the surviving objects in Figure \ref{fig:tamsobre}. Fitting this histogram to a Gaussian curve, we find that the mean radius of these bodies is $18.2\pm 2.6$ km and the standard deviation is $10.1$ km.

\section{Conclusions}
\label{sec:con}

In this work we pose an alternative method for planetesimals being captured in co--orbital resonance with a forming giant planet during its migration phase embed in a gaseous disk. We studied this process in different scenarios and we found that even if the growth of the planet is not considered as it does in \citet{1998A&A...339..278M}, the capture has a not negligible co--orbital capture probability, consistent with previous works \citep{2008A&A...481..519C}. 

The dependence on the migration rate is also analysed. We found that a faster migrating embryo has more probability to capture planetesimals in co--orbital resonance due to a reduced difference between migration timescales of both bodies. Nevertheless, the planetesimals are more easily lost in this case owing to a larger speed relative to the gas, increasing the gas drag force.

Notwithstanding, as other mechanism had been proposed as being responsible for the co--orbital capture during the same stage of planetary formation, we also investigate about the fate of bodies captured in the 1:1 resonance with a giant planet under the effect of the gas. We find that the gap carved in the gas disk by the planet plus the presence of a circumplanetary disk enhance the survival probability, making possible to have bodies in the co--orbital region prior to the planetary instability. Also, the escaping bodies are mostly pushed to the outer parts of the system.

The size distribution of the bodies surviving the migration and the friction with the gas is not uniform. Instead, if we fit the radius of the surviving objects with a Gaussian function, we find that the mean radius is $18.2$ km. Using the mean albedo found with Spitzer data by \citet{2009AJ....138..240F}, this leads to a mean absolute magnitude of $H=11.6$. This value is smaller than the one observed at the present \citep{2014SoSyR..48..139S}, but considering the subsequent collisional evolution these objects may have had undergo, this is a coherent result. 

It is worth noting that the search for possible solutions to the problems of the method of capture invoked in Section \ref{sub:sur} is out of the scope of this work. We are not posing that this population of bodies is exactly the same as the present Trojans, but the existence of bodies in the co--orbital region before the instability proposed in \citet{2009A&A...507.1041M}, could be the origin of at least a part of the population observed nowadays, presumably those with lower orbital inclinations. Previous works support this point of view, for example \citet{2007MNRAS.377.1393S} find that the color distribution for Trojans with $i<10^{\circ}$ is statistically different for those with $i>10^{\circ}$, which may indicate a different origin.

In conclusion, despite only having studied a few cases, we have found encouraging results which are consistent with previous works and support our hypothesis. In addition to increasing the number of experiments, future work will focus on two different aspects. In the first place the mass growth, which has proved to be an important effect in the capture process, will be taken into account, in order to determine its relevance not only in the capture process, but also in extending the time the planetesimals stay in the co--orbital region. Secondly, we will focus on investigation of the effects that planetary instability, including encounters with a giant planet, have over the captured objects, to test if they can survive this scenario or get lost during it.

\section*{Acknowledgements}

This work was supported by CSIC Grupos I+D program through project 831725 - Planetary Sciences and PEDECIBA Fisica. PL would like to acknowledge financial support by ANII under National Postgraduate Studies Grant POS\_NAC\_2015\_1\_109492. The authors thank the anonymous referees for suggestions that improved this article.



\section*{References}
  \bibliographystyle{elsarticle-harv} 
  \bibliography{biblio}

%\begin{thebibliography}{00}


%\bibitem{}

%\end{thebibliography}
\end{document}